\documentclass[conference]{IEEEtran}
\IEEEoverridecommandlockouts
\usepackage{cite}
\usepackage{amsmath,amssymb,amsfonts}
\usepackage{algorithmic}
\usepackage{graphicx}
\usepackage{url}
\usepackage{tikz}
\usetikzlibrary{arrows.meta, positioning, fit, backgrounds, decorations.pathreplacing, calc}
\usepackage[T1]{fontenc}
\usepackage{listings}

\usepackage{booktabs}
\newcommand{\lhl}[1]{%
  {\setlength{\fboxsep}{0pt}%
   \colorbox{yellow!45}{\scriptsize\ttfamily\upshape\strut #1}}}

\newcommand{\lhla}[1]{%
  {\setlength{\fboxsep}{0pt}%
   \colorbox{red!22}{\scriptsize\ttfamily\upshape\strut #1}}}
   
\usepackage{xcolor}
\usepackage{libertinust1math} 
\usepackage{textcomp}

\def\BibTeX{{\rm B\kern-.05em{\sc i\kern-.025em b}\kern-.08em
    T\kern-.1667em\lower.7ex\hbox{E}\kern-.125emX}}
\begin{document}

\title{A gem5-based Simulation Framework for Computing-in-DRAM}

\author{%
\IEEEauthorblockN{%
Alexander Kusnezoff\textsuperscript{1},
Jo\~ao Paulo C. de Lima\textsuperscript{3,*},
Jeronimo Castrillon\textsuperscript{1,2},
Asif Ali Khan\textsuperscript{4,*}%
}
\IEEEauthorblockA{%
\textsuperscript{1}\textit{TU Dresden},
\textsuperscript{2}\textit{ScaDS.AI},
\textsuperscript{3}\textit{Friedrich-Alexander-Universität Erlangen-Nürnberg} \\
\textsuperscript{4}\textit{Paderborn University, Germany} \\
alexander.kusnezoff@mailbox.tu-dresden.de, joao.lima@fau.de, jeronimo.castrillon@tu-dresden.de\\ asif.khan@uni-paderborn.de%
}%
}


\newboolean{showcomments}
\setboolean{showcomments}{true}
\makeatletter
\newcommand{\mynote}[3]{%
  \ifthenelse{\boolean{showcomments}}{%
   \fbox{\bfseries\sffamily\scriptsize#1}%
   {\small$\blacktriangleright$\textsf{\emph{\color{#3}{#2}}}$\blacktriangleleft$}}%
  {%
   \@bsphack
   \@esphack
  }%
}
\makeatother

\definecolor{ambit-cmd}{RGB}{180,80,0}      
\definecolor{ambit-const}{RGB}{0,100,160}   
\definecolor{ambit-tmp}{RGB}{80,80,80}      
\definecolor{ambit-comment}{RGB}{34, 139, 34}  

\lstdefinelanguage{AmbitASM}{
    morekeywords=[1]{AAP, AP},
    morekeywords=[2]{C0, C1, DCC0, DCC1, DCC},
    morekeywords=[3]{T0, T1, T2, T3},
    keywordstyle=[1]\color{ambit-cmd}\bfseries\upshape,
    keywordstyle=[2]\color{ambit-const}\bfseries\upshape,
    keywordstyle=[3]\color{ambit-tmp}\upshape,
    comment=[l]{//},
    commentstyle=\color{ambit-comment}\upshape\footnotesize,
    sensitive=true,
}

\lstdefinestyle{ambitstyle}{
    language=AmbitASM,
    basicstyle=\ttfamily\footnotesize\upshape,
    backgroundcolor={},
    frame=none,
    numbers=left,
    numberstyle=\tiny\color{gray},
    numbersep=6pt,
    xleftmargin=1.5em,
    columns=fullflexible,
    keepspaces=true,
    showstringspaces=false,
    breaklines=false,
    aboveskip=2pt,
    belowskip=2pt,
    morecomment=[l]{//},
    escapeinside={(*@}{@*)},
}

\definecolor{cpp-kw}{RGB}{0,0,160}

\lstdefinestyle{cppstyle}{
    language=C++,
    basicstyle=\ttfamily\footnotesize\upshape,
    keywordstyle=\color{cpp-kw}\bfseries,
    morekeywords=[2]{int8_t,int16_t,int32_t,int64_t,
                     uint8_t,uint16_t,uint32_t,uint64_t,size_t},
    keywordstyle=[2]\color{ambit-const}\bfseries,
    morekeywords=[3]{pimMalloc,pimFree,
                     rowand,rowor,rownot,rowxor,
                     rowadd,rowsub,rowmult,rowdiv,
                     rowmin,rowmax,rowequal,rowgreater,rowgreater_equal,
                     rowif_else,rowabs,rowbitcount},
    keywordstyle=[3]\color{ambit-cmd}\bfseries,
    commentstyle=\color{ambit-comment}\upshape\footnotesize,
    stringstyle=\color{ambit-const},
    backgroundcolor={},
    frame=none,
    numbers=left,
    numberstyle=\tiny\color{gray},
    numbersep=6pt,
    xleftmargin=1.5em,
    columns=fullflexible,
    keepspaces=true,
    showstringspaces=false,
    breaklines=true,
    aboveskip=2pt,
    belowskip=2pt,
}

\lstset{
    basicstyle=\ttfamily\small,
    backgroundcolor=\color{gray!10},
    frame=single,
    commentstyle=\color{gray},
}     

\definecolor{asparagus}{rgb}{0.53, 0.66, 0.42}
\newcommand{\jc}[1]{\mynote{Jeronimo}{#1}{asparagus}}
\newcommand{\ak}[1]{\mynote{AK}{#1}{blue}}
\newcommand{\alk}[1]{\mynote{Alex}{#1}{violet}}
\newcommand{\jp}[1]{\mynote{JP}{#1}{purple}}

\maketitle

\begingroup
\renewcommand\thefootnote{*}
\footnotetext{Work done while the authors were at TU Dresden.}
\endgroup

\begin{abstract}
Computing-in-Memory using DRAM (CIMD) has demonstrated substantial energy and throughput gains for memory-bound workloads consisting of bulk-bitwise operations, by performing computation directly within DRAM
subarrays. Realizing CIMD, however, requires a redesign of the memory
controller and careful mapping of operands onto the memory arrays.
Presently, accurate FPGA-based testbeds exist, but they are costly and labor-intensive, while open-source simulators are largely trace-based and cannot execute full application runs. The only full-application CIMD simulator available, included
with MIMDRAM\cite{oliveira2024mimdram}, is built on an outdated version of gem5 and its toolchains. We
present gem5-CIMD, a full-application CIMD simulation framework
built on the latest version of gem5. We extend the simulated instruction set from four
bitwise operations to sixteen CIM instructions spanning arithmetic,
relational, conditional, and utility operations across a range of data types and bitwidths (4 to 64 bits). A complete memory-management stack comprising multi-level huge-page support, a CIM-aware allocator, and a subarray-aware address-mapping scheme ensures that CIM operands are co-located in the same DRAM subarray. In addition, gem5-CIMD is aware of the DRAM refresh interval and
faithfully models and schedules the mandatory refresh operations when needed. We provide a CIM standard library intended as a compiler
target and evaluate gem5-CIMD on multiple case studies including an end-to-end KNN workload. The simulator sources, including the evaluated workloads, are available
at \url{https://github.com/tud-ccc/gem5-CIMD}.
\end{abstract}


\section{Introduction}
\label{sec:intro}
Data movement, not computation, dominates the energy and latency budget of modern memory-bound workloads~\cite{boroumand2018google}. Across machine learning, database, and graph-processing applications, moving operands between memory and the processor is the principal bottleneck, accounting for over 62\% of total system energy consumption~\cite{boroumand2018google, mutlu2024memory, lee2024pim}. This gap between compute throughput and memory bandwidth continues to widen as Moore's Law slows down. 
Processing-in-memory (PIM) addresses this directly by performing computation within, or close to, the memory that stores the data, thus reducing or fully eliminating data transfers~\cite{khan2024landscape, mutlu2024memory}. 

PIM has been demonstrated across various memory technologies and computational paradigms, targeting a broad range of application domains~\cite{khan2024landscape}. Among them, computing-in-memory using DRAM (CIMD)\footnote{Also referred to as Processing-using-DRAM (PuD).} is especially attractive for applications requiring bulk-bitwise logic and arithmetic operations. By simultaneously \emph{activating} multiple rows, CIMD repurposes the analog charge-sharing behavior of DRAM sense amplifiers to implement basic logic operations directly within the array~\cite{seshadri2017ambit, yuksel2024functionally, shin2024processing}, from which the full set of bitwise and arithmetic operations can be derived~\cite{hajinazar2021simdram}. For instance, in the Ambit scheme~\cite{seshadri2017ambit}, simultaneously activating three rows causes the charge accumulated on each column's bitline to reflect the \textit{bitwise majority } of the three stored values and a subsequent \textit{precharge} operation closes all rows and writes the majority result back to each of them. This strictly  requires operands to be column-aligned within the same subarray. Other architectures such as SIMDRAM~\cite{hajinazar2021simdram}, MIMDRAM~\cite{oliveira2024mimdram}, PRADA~\cite{shin2024processing}, and FCDRAM~\cite{yuksel2024functionally} demonstrate that a wide range of operations can be expressed as sequences of such \textit{activate} and \textit{precharge} (\texttt{AP}) primitives, collectively referred to as microprograms~\cite{hajinazar2021simdram}, and executed \textit{in-place}.

Leveraging CIMD in practice requires error-prone low-level programming or novel compilers. 
There is a growing body of work for the latter, where high-level compilation frameworks take an application kernel as input and generate its corresponding microprograms~\cite{guo2025pimsynth, peng2023chopper, khan2024cinm, shin2023pimflow, yang2025tensor}. For validation, these works either rely on ad-hoc, in-house simulators or on trace-driven memory-only simulators. The few available hardware testbeds, such as PiDRAM~\cite{olgun2022pidram}, run on FPGA prototypes that are accurate but costly to acquire and laborious to configure, placing them out of reach for most researchers.

Simulation is the natural alternative, offering fine-grained control over system configuration, inexpensive design-space exploration, and flexible integration into compiler pipelines. Existing CIMD simulators, however, fall short. Ramulator~\cite{ramulatorpim}- and NVMain-based simulators~\cite{nvmainpim} operate on memory traces and cannot execute full application runs, capturing the DRAM device in isolation while ignoring the full-system aspects. To the best of our knowledge, the only simulator capable of running complete CIMD applications is the one released with MIMDRAM~\cite{oliveira2024mimdram}.\footnote{\url{https://github.com/CMU-SAFARI/MIMDRAM}}
However, this simulator is built on an outdated version of gem5 (2015), requiring outdated Python and GCC toolchains, and has been 
isolated from nearly a decade of upstream advances in memory-controller modeling, cache-coherence support, and system emulation. 
Beyond the fundamental \texttt{AP} and \texttt{AAP} operations, the simulator exposes only four bitwise primitives (AND, OR, NOT, XOR) and lacks the memory-management infrastructure needed to guarantee subarray co-location of operands. Arithmetic operations such as addition or multiplication can only be expressed by manually decomposing them into raw \texttt{AP} and \texttt{AAP} sequences, thus requiring the programmer to implement carry propagation and manage temporary DRAM rows explicitly with no ISA-level abstraction for arithmetic or relational operations. 

\begin{table}[t]
\caption{Feature comparison of gem5-CIMD vs.\ MIMDRAM.}
\label{tab:mimdram-cmp}
\centering
\footnotesize
\renewcommand{\arraystretch}{1.25}
\begin{tabular}{lcc}
\toprule
\textbf{Feature} & \textbf{MIMDRAM} & \textbf{gem5-CIMD} \\
\midrule
gem5 version                       & 2015-era      & v25 (2025)  \\
Supported CIM ISA                  & 4 ops         & 16 ops      \\
Same-subarray operand check        & $\times$    & \checkmark  \\
CIMD-specific memory management & $\times$      & \checkmark  \\
\texttt{AP}/\texttt{AAP} statistics counters         & $\times$      & \checkmark  \\
\bottomrule
\end{tabular}
\end{table}

This paper proposes gem5-CIMD, a CIMD simulation framework built on the latest gem5 (v25)~\cite{gem5} that addresses the above-mentioned limitations. Table~\ref{tab:mimdram-cmp} compares gem5-CIMD with the baseline MIMDRAM implementation, highlighting the newly added features.
Concretely, our contributions are as follows:  

\noindent\textbf{Extended instruction set:} gem5-CIMD extends the instruction set from four bitwise operations to sixteen CIMD instructions, spanning arithmetic, relational, conditional, and utility operations at bitwidths of 4, 8, 16, 32, and 64 bits, each backed by a precomputed Ambit~\cite{seshadri2017ambit} microprogram per supported bitwidth, executed with cycle-accurate DRAM timing.

\noindent\textbf{Memory-management support:} We implement multi-level huge-page table support in gem5's syscall-emulation mode, a CIM-aware memory allocator, and a subarray-aware address-mapping scheme that guarantees operands sharing a computation are placed in the same DRAM subarray.

\noindent\textbf{Realistic refresh modeling:} 
 Microprograms in CIMD quickly become longer than a typical DRAM refresh interval, e.g., a 64-bit addition expands to 597 \texttt{AP}/\texttt{AAP} commands. 
 We observed that this is an aspect that has been neglected so far in existing simulators. 
 By omitting mandatory interleaved refresh operations, such simulators yield optimistic performance estimates.  
 To account for this, gem5-CIMD models intermediate refreshes explicitly.

\noindent\textbf{Library support:} We expose the full instruction set as a C/C++ standard library intended as a compiler target. We validate gem5-CIMD against an analytical command model and the MIMDRAM simulator, and evaluate multiple case studies including bitwise logic, arithmetic, relational, and conditional operations as well as an end-to-end KNN classifier as a qualitative proof of concept. 

The remainder of this paper is organized as follows.
Section~\ref{sec:background} provides a short background on CIMD and the gem5 simulator. Section~\ref{sec:framework} describes the gem5-CIMD framework.
Section~\ref{sec:eval} presents the evaluation, and
Section~\ref{sec:conclusion} concludes the paper.

\section{Background and Related Work}
\label{sec:background}

\subsection{CIM using DRAM}
\label{subsec:pud}
A DRAM device is organized hierarchically into channels, ranks, banks, and 
subarrays. Each subarray is a two-dimensional grid of cells, where each cell 
consists of an access transistor and a storage capacitor holding one bit of 
data. Cells are arranged at the intersections of wordlines (rows) and bitlines 
(columns), with a sense amplifier connected to each bitline. A memory 
controller (MC) translates each physical address into a (rank, bank, subarray, 
row, column) tuple and drives the device with a small command set. A typical 
read access proceeds as follows: an activate (\texttt{ACT}) command raises the target wordline, connecting the storage capacitors of the selected row to their respective bitlines; the stored charge shifts the precharged bitline voltage ($0.5\,V_{DD}$) by a small 
$\pm\delta$, which the sense amplifiers detect and amplify to full $V_{DD}$ or 
$GND$, latching the row into the row buffer. Read/write (\texttt{RD}/\texttt{WR}) commands 
then transfer data between the row buffer and the memory bus. Finally, a precharge (\texttt{PRE}) command closes the row and restores all bitlines to $0.5\,V_{DD}$ in 
preparation for the next access. Separately, \texttt{REF} commands must be 
issued periodically at refresh interval $t_{REFI}$ to restore charge 
lost to leakage; otherwise, the stored data gets corrupted. All commands are governed by 
timing constraints, including $t_{RAS}$, $t_{RCD}$, and $t_{RC}$, that bound 
the minimum time between operations and guarantee correct device behavior.

CIMD repurposes the default behavior of DRAM either by deliberately violating 
timing constraints~\cite{yuksel2024functionally, shin2024processing}, or by 
combining timing violations with DRAM command 
reordering~\cite{seshadri2013rowclone, seshadri2017ambit}, to perform 
computation in-place. RowClone~\cite{seshadri2013rowclone} copies a row in 
place by issuing two back-to-back \texttt{ACT}s to the same subarray followed by a \texttt{PRE} 
($\texttt{ACT}\langle src\rangle$, $\texttt{ACT}\langle dst\rangle$, 
\texttt{PRE}, abbreviated \texttt{AAP}), so the source row's values overwrite the 
destination cells without leaving the array. For computations, the exact compute mechanism, i.e., DRAM commands and their order, 
depends on the CIMD architecture. For instance, in Ambit~\cite{seshadri2017ambit}, activating 
three rows simultaneously followed by a precharge operation (\texttt{AP})  
makes the sense amplifiers resolve the bitwise majority of the three operands, 
yielding $\mathrm{MAJ}(a,b,c)$. Pinning one input to $0$ or $1$ gives 
\texttt{AND} and \texttt{OR}, respectively, and a dual-contact cell (DCC) 
provides \texttt{NOT}. FCDRAM~\cite{yuksel2024functionally} and 
PRADA~\cite{shin2024processing} achieve the same effect on unmodified commodity 
DRAM purely through timing-violating \texttt{ACT}/\texttt{PRE} sequences. 
Because all participating rows must reside in the same subarray (or neighboring subarrays, as in FCDRAM) and TRA is 
destructive, operands are typically first RowClone-copied into reserved compute rows before each operation. Higher-level and more complex operations are represented by decomposing them into longer sequences of \texttt{AP} and \texttt{AAP} commands, which are usually referred to as \emph{microprograms}. 

Figure~\ref{fig:uProgram} shows the microprogram for a logical \texttt{AND} between two rows. 
Deconstructing large operations into such fundamental logic operations can thus quickly lead to large sequences of memory commands. 
As an example, a 64-bit addition expands to 
597 \texttt{AP}/\texttt{AAP} commands, which is why faithfully accounting for 
interleaved \texttt{REF} commands is essential for correct timing simulation.

\begin{figure}[t]
\centering
\begin{lstlisting}[style=ambitstyle]
// AND(a,b) = MAJ(a,b,0) -> c
AAP  a,  T0       // a  -> T0
AAP  b,  T1       // b  -> T1
AAP  C0, T2       // 0  -> T2
AP   T0, T1, T2   // MAJ(a,b,0) = AND(a,b)
AAP  T0, c        // result -> c
// NOT(a) -> c
AAP  a,   DCC0     // a->DCC0, NOT(a)->DCC1
AAP  DCC1, c      // NOT(a) -> c
\end{lstlisting}

\caption{Ambit microprograms for \texttt{AND} and \texttt{NOT}~\cite{seshadri2017ambit}.
         \texttt{T0}--\texttt{T2}: B-group compute rows;
         \texttt{C0}: constant-zero row;
         \texttt{DCC}: dual-contact cell (stores $x$ and $\lnot x$).
         All rows must reside in the same subarray.}
\label{fig:uProgram}
\end{figure}

\subsection{The gem5 Simulator}
\label{subsec:gem-sim}
gem5~\cite{gem5} is a modular, cycle-accurate simulator widely used in computer
architecture research. It executes unmodified binaries and reports detailed
per-component statistics, including a per-rank DRAM energy breakdown across
\texttt{ACT}, \texttt{PRE}, read, and write commands. Target systems are described in
Python configuration scripts that expose CPU, cache, and memory parameters such
as DDR timings, banks per rank, capacity, and the address-mapping scheme. gem5
offers two execution modes: Full-System (FS) mode boots a complete OS image,
while Syscall-Emulation (SE) mode intercepts system calls with lightweight stubs
and replaces OS structures such as the page table with its own
\texttt{EmulationPageTable}, simplifying setup and reducing runtime.

The memory subsystem is structured as a class hierarchy that gem5-CIMD extends.
\texttt{AbstractMemory} is the base class for all memory devices and 
handles functional memory accesses; 
\texttt{MemInterface} holds the address-mapping policy and shared parameters; and
\texttt{DRAMInterface} specializes it with DRAM timing parameters and helpers
such as \texttt{activateBank()} and \texttt{prechargeBank()} that issue the
underlying \texttt{ACT}/\texttt{PRE} commands. \texttt{MemCtrl} schedules
requests onto the interface. This layer is where the CIM command primitives and
microprogram execution are introduced.

\subsection{Related Work}
\label{subsec:relatedwork}
Validating the output of a CIMD compiler requires either a hardware testbed or a
simulator. FPGA-based platforms such as PiDRAM~\cite{olgun2022pidram} and
DRAM-Bender~\cite{olgun2023drambender} expose a DRAM interface for issuing
arbitrary, timing-relaxed command sequences and reproduce real device behavior,
but they are costly to acquire and laborious to configure, placing them out of
reach for most compiler developers. Software simulators lower this barrier, yet
existing options are limited: Ramulator-PIM~\cite{ramulatorpim} and
NVMain-PIM~\cite{nvmainpim} model the DRAM device in isolation, operating
on memory-command traces and ignoring the non-memory instructions of a real
program, so they cannot run full applications.

To our knowledge, MIMDRAM~\cite{oliveira2024mimdram} is the only CIMD simulator
that executes complete binaries. Its gem5 base, however, dates to 2015, predating
formal gem5 versioning and isolating it from a decade of upstream
advances, and it exposes only four bitwise primitives (\texttt{AND}, \texttt{OR},
\texttt{NOT}, \texttt{XOR}). Its public release also omits several mechanisms
described in the paper that are essential for executing real CIMD workloads, such as 
same-subarray operand checks, a subarray-aware address-mapping scheme (the
default bank-interleaving scatters the elements of an array across banks), a
reserved address space with CIM-aware allocation, and statistics for executed CIM
operations. gem5-CIMD implements all these features using the latest gem5, providing a
full-application platform for validating hand-written or compiler-generated CIMD code.

\section{The gem5-CIMD Framework}
\label{sec:framework}
gem5-CIMD ports and extends the simulator released with MIMDRAM~\cite{oliveira2024mimdram}. The porting itself required non-trivial changes to the code base, including rewriting most of the API function calls (e.g., \texttt{DRAMPacket}$\to$\texttt{MemPacket},
\texttt{DRAMCtrl}$\to$\texttt{MemInterface}+\texttt{MemCtrl} etc.). Beyond the port, we add new features in gem5-CIMD that are essential for modern CIMD systems. 
These include an extended 16-instruction ISA, subarray-level operand placement, memory management, and DRAM-refresh interleaving during long microprograms.

\begin{figure*}
    \centering
\includegraphics[width=0.75\linewidth]{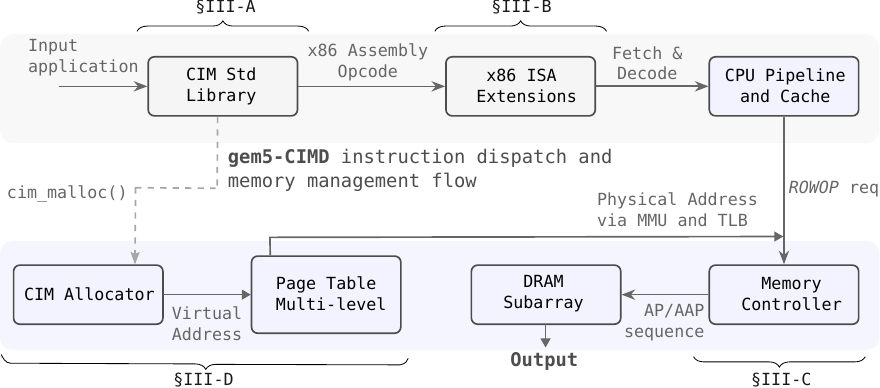}
    \caption{gem5-CIMD framework overview. The execution flow (top, left to
right) traces a CIM instruction from the input application through the
ISA and cache hierarchy. The bottom part shows
the memory management including huge-page allocation and address translation. The memory controller decodes instructions into \texttt{AP}/\texttt{AAP} sequences and uses the physical address to execute commands in the DRAM subarray.}
\label{fig:overview}
\end{figure*}

Figure~\ref{fig:overview} shows an overview of gem5-CIMD. An input program (either hand-written or compiler-generated code) invokes the CIM standard library (\S\ref{subsec:stdlib}), which emits custom x86 opcodes via the ISA extension (\S\ref{subsec:isa}).
The CPU pipeline fetches and decodes these as memory-referencing instructions,
forwards them through the cache hierarchy with cache-line invalidation, and
delivers a \texttt{ROWOP} request to the memory controller, which decodes it
into \texttt{AP}/\texttt{AAP} sequences executed in the DRAM subarray (\S\ref{subsec:mc}).
Orthogonally, \texttt{cim\_malloc()} allocates operands through the memory
management stack (\S\ref{subsec:memstack}), guaranteeing that all operands
participating in a CIM operation are co-located in the same DRAM subarray.

\subsection{CIM Standard Library}
\label{subsec:stdlib}
The CIM standard library is the primary interface between user code and the
gem5-CIMD framework. It exposes two layers: a memory-management layer
with \texttt{cim\_malloc} and \texttt{cim\_free}, and a computation layer providing
C/C++ wrappers around the sixteen CIM instructions. Each wrapper emits
the corresponding custom x86 opcode via inline assembly, passing operand
addresses through a fixed register convention: \texttt{RDI} = destination,
\texttt{RSI} = first source, \texttt{RDX} = second source, \texttt{RCX} =
element count, \texttt{R8} = element bitwidth, and \texttt{R9} = mask (\texttt{rowif\_else} only).\footnote{As one would expect, not all registers are used in every CIM instruction.} The bitwidth parameter
enables sub-byte operand support (e.g., \texttt{INT4}) without any changes
to the underlying hardware model. Table~\ref{tab:stdlib} lists all
provided routines.

\begin{table}[tbh]
\caption{CIM standard library routines.}
\label{tab:stdlib}
\centering
\footnotesize
\renewcommand{\arraystretch}{1.25}
\begin{tabular}{lp{5.8cm}}
\toprule
\textbf{Category} & \textbf{Routines} \\
\midrule
Memory      & \texttt{cim\_malloc}, \texttt{cim\_free} \\
Bitwise     & \texttt{rowand}, \texttt{rowor}, \texttt{rownot}, \texttt{rowxor} \\
Arithmetic  & \texttt{rowadd}, \texttt{rowsub}, \texttt{rowmult}, \texttt{rowdiv} \\
Relational  & \texttt{rowmin}, \texttt{rowmax}, \texttt{rowequal}, \texttt{rowgreater},
              \texttt{rowgreater\_equal} \\
Conditional & \texttt{rowif\_else} \\
Utility     & \texttt{rowabs}, \texttt{rowbitcount} \\
\bottomrule
\end{tabular}
\vspace{-0.5cm}
\end{table}
     
Figure~\ref{fig:stdlib-usage} compares the two representations of the same element-wise \texttt{INT4} addition kernel.
The yellow regions show how a single
\texttt{rowadd} call in~(a) requires the programmer to manually write the entire carry-ripple decomposition in~(b).
MIMDRAM does provide an LLVM pass handling \texttt{AAP\_VECTORS}/\texttt{AP\_VECTOR} macros that iterate over bit positions rather than fully unrolling all \texttt{AP}/\texttt{AAP} commands, but the programmer must still express every step of the carry logic explicitly.
This burden grows with operation complexity, e.g., a 32-bit multiplication still reduces to one library call in~(a) yet expands to over 16K~\texttt{AP}/\texttt{AAP}
commands in~(b), making manual authoring both error-prone and impractical.
For complete application workloads such as the KNN classifier evaluated in
Section~\ref{sec:eval}, expressing the computation as raw \texttt{AP}/\texttt{AAP} sequences and manually handling their reduction across columns 
becomes infeasible altogether.

\subsection{ISA Extension}
\label{subsec:isa}
gem5-CIMD extends the x86 ISA with sixteen new instructions, whose opcodes were relocated from those used by MIMDRAM to remain compatible with modern gem5, covering bitwise, arithmetic, relational, conditional, and utility operations at bitwidths 4, 8, 16, 32, and 64~bits backed by a library of 16 (instructions) $\times$ 5 (bitwidths) = 80 precomputed Ambit microprograms (see \S\ref{subsec:mc}). When the CPU decodes a CIM instruction, it creates a memory request with a \texttt{ROWOP} flag and attaches a \texttt{RowOpPayload} carrying the
operation type, operand addresses, element count, and bitwidth. For
operations with multiple operands, \texttt{splitRowOp()} generates
independent sub-requests for each address, each undergoing a separate TLB
translation. At the cache level, all cache lines overlapping with the
destination and source rows are written back and invalidated, and the
request is forwarded only after these write-backs (and retried if they do
not fit into the write buffer), ensuring that the operation computes on the
latest data and that subsequent CPU reads get the correct result rather
than a stale cached value.

\subsection{Memory Controller and Refresh Handling}
\label{subsec:mc}
The memory controller (MC) is extended at two levels. \texttt{MemCtrl} is
extended to decode incoming \texttt{ROWOP} requests. It calls
\texttt{decodePacket()} for each operand address to obtain the precise DRAM
address (rank, bank, subarray, row) and asserts that all operands map
to the same bank, rank, and subarray - a hard requirement of Ambit-based
CIM. The bus state machine is modified to prioritize pending CIM
operations before resuming regular read/write accesses. Statistics
counters track the number of TRA commands (\texttt{AP}s, see \S\ref{subsec:pud}) and RowClone operations (\texttt{AAP}s) executed, enabling detailed performance analysis.

\begin{figure}[t]
\centering
\begin{lstlisting}[style=cppstyle, escapechar=§,
  basicstyle=\ttfamily\scriptsize\upshape,
  commentstyle=\color{ambit-comment}\upshape\scriptsize]
#include "cim_core.h"

int main() {
  /* int8_t: min C type; BITS=4 sets CIM op width */
  const size_t N = 1024, BITS = 4;
  int8_t *a   = §\lhla{cim\_malloc(N*sizeof(int8\_t), 0);}§
  /* same for *b and *dst */
  
  for (size_t i = 0; i < N; ++i) // Initialization
    { a[i] = (int8_t)i;  b[i] = 2*(int8_t)i; }
  §\lhl{rowadd(dst, a, b, N, BITS);}§ /* dst[i]=a[i]+b[i] */
  /* display / consume output */
  cim_free(a);  cim_free(b);  cim_free(dst);
  return 0;
}
\end{lstlisting}
\vspace{-3pt}
\centerline{\footnotesize (a) gem5-CIMD: C/C++ standard library}
\vspace{4pt}
\begin{lstlisting}[style=cppstyle, escapechar=§,
  basicstyle=\ttfamily\scriptsize\upshape,
  commentstyle=\color{ambit-comment}\upshape\scriptsize]
// No cim_core.h; ADD not in MIMDRAM ISA.

int main() {
  const size_t N = 1024, BITS = 4;
  /* malloc: no automatic subarray co-location */
  int8_t *a   = §\lhla{malloc(N*sizeof(int8\_t));}§
  /* same for *b and *dst */
  
  for (size_t i = 0; i < N; ++i) // Initialization
    { a[i] = (int8_t)i;  b[i] = 2*(int8_t)i; }
  
  §\lhl{AAP (a,  T0)}§      // a      -> T0
  §\lhl{AAP (b,  T1)}§      // b      -> T1
  §\lhl{AAP (C0, T2}§      // cin=0  -> T2  (carry for bit 0)
  §\lhl{AP  (T0, T1, T2}§  // carry  = MAJ(a,b,cin)
  §\lhl{AAP (T0, DCC0}§    // carry  -> DCC0/DCC1
  §\lhl{AAP (a,  T0}§      // reset  T0 = a
  §\lhl{AAP (b,  T1}§      // reset  T1 = b
  §\lhl{AP  (T0, T1, DCC1)}§ // sum   = XOR(a,b,cin)
  §\lhl{AAP (T0, dst)}§     // sum    -> dst
  // carry[i] feeds bit i+1; full INT4 add: 41 AP/AAPs
  /* display / consume output */
  free(a);  free(b);  free(dst);
  return 0;
}
\end{lstlisting}
\vspace{-3pt}
\centerline{\footnotesize (b) MIMDRAM: \texttt{AP}/\texttt{AAP} microprogram~\cite{seshadri2017ambit,hajinazar2021simdram}}
\caption{The same element-wise \texttt{INT4} addition in gem5-CIMD (a) and MIMDRAM (b). \colorbox{red!22}{Allocation}: \texttt{cim\_malloc}
  guarantees subarray co-location automatically;
  \texttt{malloc} does not. \colorbox{yellow!45}{Computation}: one \texttt{rowadd} call vs.\ the full carry-ripple   decomposition in~(b); MIMDRAM's \texttt{AAP\_VECTORS}/\texttt{AP\_VECTOR}
  macros loop over bit positions but the programmer writes the carry logic manually.
  \texttt{T0}--\texttt{T2}: B-Group rows; \texttt{C0}: constant-zero row;
  \texttt{DCC}: dual-contact cell ($x$ and $\lnot x$).}
\label{fig:stdlib-usage}
\vspace{-0.4cm}
\end{figure}

\texttt{DRAMInterface} is extended to support two new primitives. 
\texttt{apBank()} performs a TRA (AP) operation, computing
$\mathrm{MAJ}(a,b,c)$ across three simultaneously activated rows; and
\texttt{aapBank()} performs a RowClone (\texttt{AAP}), copying a source row to a
destination row using the overlapped-activate timing parameter
$t_{WLOV}$. For every CIM instruction, including the basic bitwise operations (AND, OR, NOT, XOR),
\texttt{executeAmbitMicroprogram()} reads the \texttt{AP}/\texttt{AAP} sequence from a library of
precomputed microprogram files, one per operation per supported bitwidth.
Users can replace or add microprograms by placing the corresponding text files in the microprogram directory, where they are picked up at runtime. This design decision is intentional, i.e., gem5-CIMD is not designed to optimize the microprogram generation. Instead, it is intended to be orthogonal to the logic synthesis part and faithfully simulate the provided microprograms. 

gem5-CIMD also faithfully models refresh operations. Specifically, in longer microprograms such as a
64-bit addition, which requires 597 \texttt{AP}/\texttt{AAP} steps, the execution time of these commands far exceeds the refresh interval
$t_{REFI} \approx 7.8\,\mu\mathrm{s}$ of a typical DRAM. Our framework monitors elapsed time during microprogram execution and inserts a synchronous \texttt{REF} command, preceded by a precharge of all banks, whenever $t_{REFI}$ has elapsed since the start of the microprogram or the previous refresh. Each refresh blocks the rank for $t_{RFC}$\footnote{The exact value of $t_{RFC}$ depends on the DRAM technology. For DDR4, $t_{RFC} = 350\,\mathrm{ns}$, according to the gem5 code base.}.
Omitting these refreshes, as in the MIMDRAM simulator, yields optimistic
latency estimates for long-running CIM operations.

\subsection{Memory Management Stack}
\label{subsec:memstack}
CIM requires all operands of a single operation to reside in the same DRAM
subarray. Standard \texttt{malloc} provides no such guarantee, and the
default bank-interleaving address mapping actively works against it by
scattering consecutive array elements across different banks. gem5-CIMD
addresses this through three coordinated mechanisms.

\subsubsection{Address Mapping} A new \texttt{RaBaSuRoCh} (Rank, Bank, Subarray,
Row, Column) mapping scheme is added to \texttt{DRAMInterface}. It
decomposes physical addresses such that consecutive bytes within a subarray row
share the same subarray, ensuring that properly aligned CIM arrays are
physically co-located. This scheme is applied uniformly to all memory
accesses to avoid mapping conflicts between CIM and non-CIM traffic.

\subsubsection{Huge-page Pool} 
A DRAM subarray typically spans several megabytes, far
exceeding a standard 4~KiB page. A reserved pool of 2~MiB huge pages is
pre-allocated at a fixed base address (\texttt{PIM\_BASE\_ADDR}), backed by
a custom \texttt{mmapPim} system call (syscall~500) emulated in gem5's SE
mode. Unlike standard \texttt{mmap}, \texttt{mmapPim} allocates pages
eagerly from \textit{contiguous physical memory}, eliminating demand-paging non-determinism. To support variable page sizes, gem5's
\texttt{EmulationPageTable} is replaced with a multi-level page table
mimicking the Linux PGD/P4D/PUD/PMD/PTE hierarchy; 2~MiB pages terminate
the walk at the PMD level, bypassing the PTE level.

\subsubsection{CIM-aware Allocator}
\texttt{cim\_malloc()} takes a size and a subarray label and provides
fine-grained allocation within the huge-page pool. The label
allows a compiler to co-locate operands that participate in the same CIM operation by assigning them identical labels: the allocator
binds each label to one allocation region and serves all co-labeled requests from it, so co-location holds by construction and is checked again by the subarray-level assertion in the
MC. Internally, each
2~MiB huge page forms one region of 256 rows of 8~KiB that lies entirely within a single DRAM subarray; allocation metadata
tracks free row ranges per region and coalesces freed blocks to reduce
fragmentation. \texttt{cim\_free()} returns blocks to the region's free list.

\section{Evaluation}
\label{sec:eval}

This section presents our experimental setup and evaluation results. We validate the timing of gem5-CIMD against an analytical command model and compare it with the MIMDRAM simulator, and demonstrate end-to-end simulation capability with a KNN classifier.

\subsection{Experimental Setup}
\label{sec:setup}
We evaluate gem5-CIMD using gem5's Syscall Emulation (SE) mode with an out-of-order \texttt{X86O3CPU} running at 4\,GHz with private L1 instruction and data caches and an L2 cache. Table~\ref{tab:gem5cfg} summarizes the simulated system configuration. The memory subsystem uses \texttt{DDR4\_2400\_8x8} with a single channel and rank. All CIM operands reside in a reserved huge-page pool mapped via \texttt{RaBaSuRoCh} to guarantee same-subarray placement.

\begin{table}[t]
\caption{gem5-CIMD system configuration.}
\label{tab:gem5cfg}
\centering
\footnotesize
\renewcommand{\arraystretch}{1.25}
\begin{tabular}{ll}
\toprule
\textbf{Component}   & \textbf{Configuration} \\
\midrule
CPU model            & X86O3CPU, 4\,GHz \\
L1 I/D caches        & 32\,KiB each, 8-way, 2 cycles \\
L2 cache             & 256\,KiB, 4-way, 20 cycles \\
Memory               & DDR4\_2400\_8x8, 1 channel, 1 rank \\
Memory capacity      & 512\,MiB \\
Address mapping      & RaBaSuRoCh \\
Simulation mode      & SE (Syscall Emulation) \\
\bottomrule
\end{tabular}
\end{table}

Our workload set covers the gem5-CIMD instruction set (Table~\ref{tab:stdlib}): four bitwise, four arithmetic, five relational, one conditional, and two utility
operations. We evaluate integer bitwidths of 8, 16, and 32 bits with array sizes of 8K, 40K, and 500K
elements. The KNN workload is used as an end-to-end application case study.

\subsection{Validation}
\label{sec:validation}
We validate the execution time of microprograms against an analytical model and the MIMDRAM simulator, the detection of misaligned operands, and the refresh modeling.

\subsubsection{Comparison with MIMDRAM and an Analytical Model}
\label{sec:hth}

Both simulators execute the same microprograms (for MIMDRAM translated into its row-operation workloads) on 1K INT32 elements on the same system (Table~\ref{tab:gem5cfg}). The analytical model takes the \texttt{AP} and \texttt{AAP} commands of each microprogram and executes them strictly in sequence, as their data dependencies through the compute rows require. Each \texttt{AP} takes $t_{RAS}+t_{RP} \approx 46\,\mathrm{ns}$ and each \texttt{AAP} $t_{RAS}+t_{WLOV}+t_{RP} \approx 50\,\mathrm{ns}$ (DDR4-2400: $t_{RAS}=32\,\mathrm{ns}$, $t_{RP}=14.16\,\mathrm{ns}$, $t_{WLOV}=4\,\mathrm{ns}$). Whenever $t_{REFI}=7.8\,\mu\mathrm{s}$ of commands have elapsed, a refresh adds $t_{RP}+t_{RFC} \approx 364\,\mathrm{ns}$ (\S\ref{sec:refresh}). The total is converted to cycles at 4\,GHz. Table~\ref{tab:hth} reports the cycles of one operation.

\begin{table}[t]
\caption{Cycles (4\,GHz) of one operation on 1K INT32 elements: analytical \texttt{AP}/\texttt{AAP} model, gem5-CIMD, and MIMDRAM ($\Delta$ relative to the model).}
\label{tab:hth}
\centering
\footnotesize
\renewcommand{\arraystretch}{1.25}
\setlength{\tabcolsep}{4pt}
\begin{tabular}{lrrrrr}
\toprule
\textbf{Kernel} & \textbf{Model} & \textbf{gem5-CIMD} & \textbf{$\Delta$} & \textbf{MIMDRAM} & \textbf{$\Delta$} \\
\midrule
add32 & 60,785    & 67,902    & $+11.7$\% & 40,791    & $-32.9$\% \\
sub32 & 203,099   & 210,215   & $+3.5$\%  & 177,530   & $-12.6$\% \\
mul32 & 3,413,936 & 3,421,362 & $+0.2$\%  & 4,205,404 & $+23.2$\% \\
eq32  & 87,033    & 94,149    & $+8.2$\%  & 61,789    & $-29.0$\% \\
\bottomrule
\end{tabular}
\vspace{-0.4cm}
\end{table}

gem5-CIMD matches the model within 0.2--12\%. The difference is a constant of about 7K cycles of CPU work in the benchmark (e.g., reading back the result), which only matters for short microprograms. MIMDRAM uses the same \texttt{AP}/\texttt{AAP} command timing, but issues every \texttt{AP} and \texttt{AAP} as a separate CPU instruction and memory request. For short microprograms, this lets it overlap commands that depend on each other, so it reports up to 33\% fewer cycles than the command sequence physically takes. For the 16K commands of the 32-bit multiplication, the per-command instruction overhead dominates and it reports 23\% more.

\subsubsection{Misalignment Detection}
\label{sec:misalign}
A fundamental requirement of Ambit-based CIM is that all operands of a single instruction must reside in the same subarray. MIMDRAM requires the programmer (or compiler) to ensure correct operand placement, and silently produces incorrect results if operands are misaligned. In contrast, gem5-CIMD enforces this with a runtime assertion, and program execution fails if this is violated.

To demonstrate this, we allocate two 32K-element INT32 arrays using standard \texttt{malloc} instead of \texttt{cim\_malloc} and execute \texttt{rowand} on both simulators. Under MIMDRAM's default
bank-interleaving address mapping, consecutive array elements are scattered across different banks. The simulator completes without error but produces bit-corrupted output. Under gem5-CIMD, the memory controller calls \texttt{decodePacket()} for each operand address and checks that all operands map to the same bank, rank, and subarray. The misaligned allocation triggers an assertion at write-queue insertion time.


\subsubsection{Refresh Handling Fidelity}
\label{sec:refresh}
Short microprograms (e.g., bitwise operations, which take 2--7 \texttt{AP}/\texttt{AAP} steps per row
independent of the bitwidth) finish well within the DRAM refresh interval ($t_{REFI} \approx 7.8\,\mu\mathrm{s}$), but longer ones must be interrupted by refreshes. MIMDRAM only refreshes between separate memory requests and never within a multi-command row operation, so its estimates for long operations issued as one request are optimistic. 

We quantify this for 64-bit addition (597 \texttt{AP}/\texttt{AAP} steps: 107~\texttt{AP}s and 490~\texttt{AAP}s). At DDR4-2400 timings, each \texttt{AP} command takes $t_{RAS} + t_{RP} \approx 46\,\mathrm{ns}$, while each \texttt{AAP} takes $t_{RAS} + t_{WLOV} + t_{RP} \approx 50\,\mathrm{ns}$, giving a total microprogram duration of $\sim$29.5\,$\mu$s---exceeding $t_{REFI}$ by a factor of
$\sim$3.8. gem5-CIMD therefore inserts $\lfloor 29.5\,\mu\mathrm{s} /
7.8\,\mu\mathrm{s} \rfloor = 3$ synchronous \texttt{REF} commands,
each consuming $t_{RP} + t_{RFC} \approx 364\,\mathrm{ns}$, adding
$\sim$1.1\,$\mu$s of overhead per operation.


\begin{figure}
    \centering
    \includegraphics[width=0.95\linewidth]{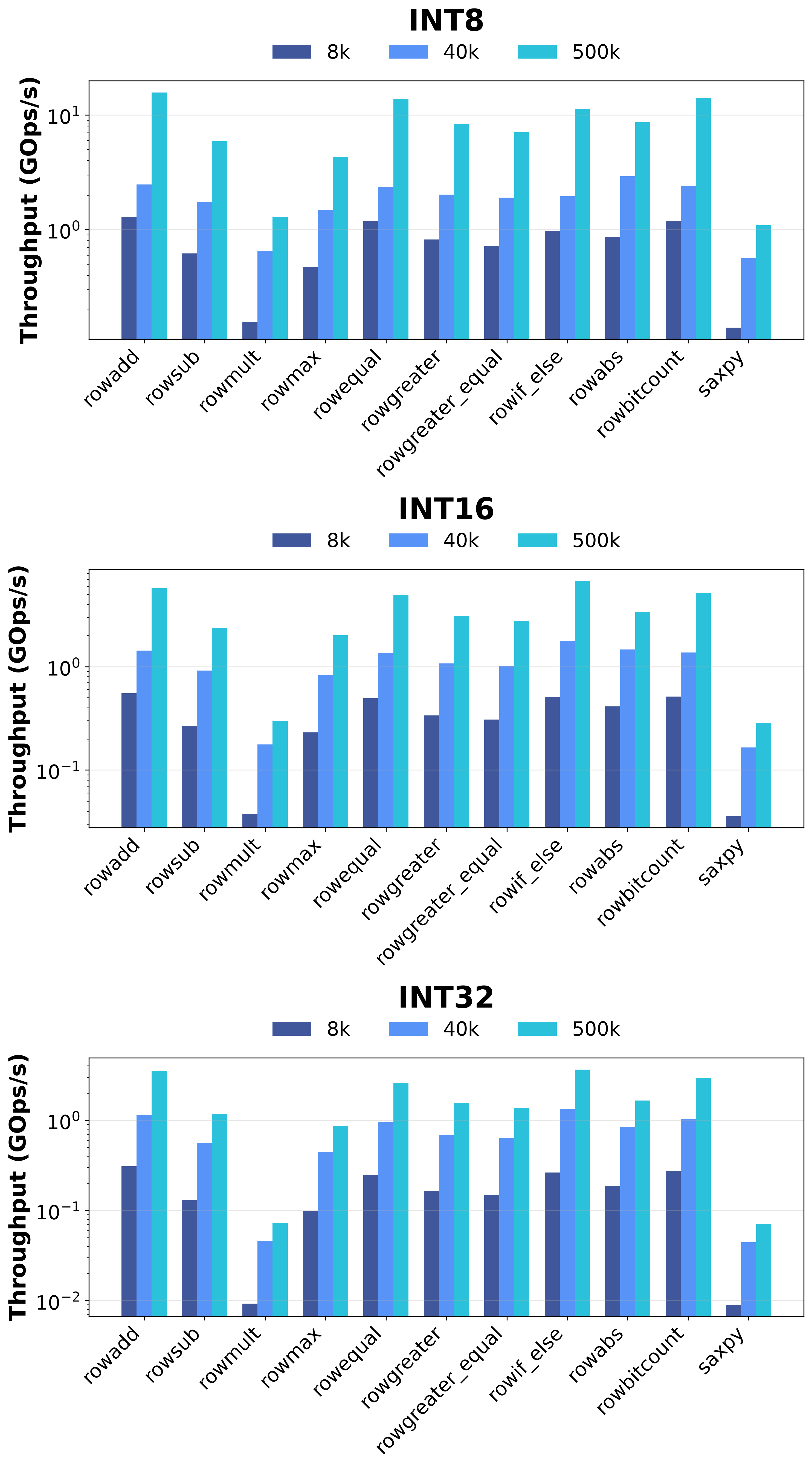}
    \caption{Throughput (GOps/s) of selected CIM workloads.}
    \label{fig:microworkloads}
\end{figure}

\subsection{Case Studies}
\label{sec:apps}
Figure~\ref{fig:microworkloads} shows the evaluation results (throughput) of 11 microworkloads for different operand dimensions (8K, 40K, 500K) and different bitwidths (8, 16, and 32 bits). 
We do not provide any comparison to CPU/GPU systems because that is not the focus of this paper---CIM gains against these architectures have been demonstrated by prior research. Our focus is instead to show that the simulation results are correct (validation in Section~\ref{sec:hth}) and that the simulator can handle running a diverse set of workloads. The measured times include writing back the operands' dirty cache lines before each operation, which is most visible at 40K elements, where the operands fit into the L2 cache. 

To demonstrate the practical applicability of gem5-CIMD, we implemented a K-Nearest Neighbors (KNN) classifier as an end-to-end workload. The KNN algorithm computes the Euclidean distance between a query point and all training samples and then
selects the $K$ nearest neighbors, whose majority vote determines the predicted class. The distance computation, which dominates runtime, is embarrassingly parallel across training samples, making it an ideal candidate for CIMD acceleration.

The implementation uses a Structure-of-Arrays (SoA) data layout where each feature column is stored as a separate CIM-allocated array. For a dataset with $F$ features and $N$ training samples, $F$ arrays of $N$ elements each are allocated in the same DRAM subarray. The squared Euclidean distance for all training samples is computed simultaneously using CIM operations:

\begin{lstlisting}[language=C, basicstyle=\ttfamily\footnotesize]
// For each feature f:
// diff = query - train
rowsub(diff, query_broadcast[f], train[f], N, BITS);
// sq = diff^2
rowmult(sq, diff, diff, N, BITS);
// dist += sq
rowadd(dist, dist, sq, N, BITS);
\end{lstlisting}

Each CIM instruction operates on all $N$ training samples in parallel within a single DRAM subarray. The host CPU handles the top-K selection and majority vote, which are inherently sequential and operate on a small number of candidates. The sources of the KNN application are available at \texttt{tests/test-progs/cim/src/combined\_knn.cpp}.

\section{Conclusion and Future Work}
\label{sec:conclusion}

We presented gem5-CIMD, a full-application CIMD simulation framework built on the latest gem5 version (v25). Building upon the MIMDRAM simulator, we ported the infrastructure to the current API and extended it by (1) increasing the number of supported instructions from four bitwise primitives to sixteen CIM instructions spanning arithmetic, relational, conditional, and utility operations, (2) implementing a memory-management stack including novel CIM-specific memory allocation and release routines that guarantee that CIM operands are co-located in the same DRAM subarray, a constraint imposed by the Ambit scheme, and (3) faithfully modeling the mandatory DRAM refresh commands, which particularly affect execution time when running long microprograms. gem5-CIMD also introduces a CIM standard library that exposes the full instruction set as a C/C++ compiler target. Its cycle counts match an analytical model of the executed AP/AAP commands within 0.2--12\%, and the differences to MIMDRAM are explained by MIMDRAM overlapping dependent commands and issuing each of them as a separate instruction. An end-to-end KNN workload demonstrates that the framework can execute complete CIM applications in syscall-emulation mode.
 
Building on this foundation, we plan to extend gem5-CIMD in the following directions. The current implementation uses gem5's SE~mode; FS~mode support would require a custom Linux kernel image with CIM system calls and a device driver managing the huge-page pool, enabling OS-level workloads and multi-process scenarios. Importantly, all the simulated results reported in this paper are for a single bank. We plan to model a more realistic DRAM, make the data layouts explicit and configurable, perform CIM in parallel in different banks, and faithfully model the data movement between banks. We also plan to extend gem5-CIMD to support other ISAs (RISC-V and ARM) and other CIMD implementations (PRADA~\cite{shin2024processing} and FCDRAM~\cite{yuksel2024functionally}).

\section*{Acknowledgment}
This work is partially funded by the AI competence center ScaDS.AI Dresden/Leipzig (01IS18026A-D), by the European Research Council (ERC) under the European Union’s Horizon Europe research and innovation programme (Consolidator Grant, grant No 101230491, project COMETH), and by the German Research Foundation (DFG) through the HetCIM project (502388442) and the TRR 404 Active 3D (528378584).

\bibliographystyle{IEEEtran}
\bibliography{references}

@article{guo2025pimsynth,
  title={{PIMsynth}: A Unified Compiler Framework for Bit-Serial Processing-In-Memory Architectures},
  author={Guo, Deyuan and Gholamrezaei, Mohammadhosein and Hofmann, Matthew and Venkat, Ashish and Zhang, Zhiru and Skadron, Kevin},
  journal={IEEE Computer Architecture Letters},
  year={2025},
  publisher={IEEE}
}

@inproceedings{shin2023pimflow,
  title={{PIMFlow}: Compiler and runtime support for {CNN} models on processing-in-memory {DRAM}},
  author={Shin, Yongwon and Park, Juseong and Cho, Sungjun and Sung, Hyojin},
  booktitle={Proceedings of the 21st ACM/IEEE International Symposium on Code Generation and Optimization},
  pages={249--262},
  year={2023}
}

@inproceedings{peng2023chopper,
  title={Chopper: A compiler infrastructure for programmable bit-serial {SIMD} processing using memory in {DRAM}},
  author={Peng, Xiangjun and Wang, Yaohua and Yang, Ming-Chang},
  booktitle={2023 IEEE International Symposium on High-Performance Computer Architecture (HPCA)},
  pages={1275--1288},
  year={2023},
  organization={IEEE}
}

@inproceedings{oliveira2024mimdram,
  title={{MIMDRAM}: An end-to-end processing-using-{DRAM} system for high-throughput, energy-efficient and programmer-transparent multiple-instruction multiple-data computing},
  author={Oliveira, Geraldo F and Olgun, Ataberk and Ya{\u{g}}l{\i}k{\c{c}}{\i}, Abdullah Giray and Bostanc{\i}, F Nisa and G{\'o}mez-Luna, Juan and Ghose, Saugata and Mutlu, Onur},
  booktitle={2024 IEEE International Symposium on High-Performance Computer Architecture (HPCA)},
  pages={186--203},
  year={2024},
  organization={IEEE}
}

@inproceedings{mutlu2024memory,
  title={Memory-Centric Computing: Recent Advances in Processing-in-{DRAM}},
  author={Mutlu, Onur and Olgun, Ataberk and Oliveira, Geraldo F and Y{\"u}ksel, {\.I}smail Emir},
  booktitle={2024 IEEE International Electron Devices Meeting (IEDM)},
  pages={1--4},
  year={2024},
  organization={IEEE}
}

@article{yang2025tensor,
  title={A Tensor Compiler for Processing-In-Memory Architectures},
  author={Yang, Peiming and Durvasula, Sankeerth and Fernandez, Ivan and Sadrosadati, Mohammad and Mutlu, Onur and Pekhimenko, Gennady and Giannoula, Christina},
  journal={arXiv preprint arXiv:2511.15503},
  year={2025}
}

@inproceedings{boroumand2018google,
  title={{Google} workloads for consumer devices: Mitigating data movement bottlenecks},
  author={Boroumand, Amirali and Ghose, Saugata and Kim, Youngsok and Ausavarungnirun, Rachata and Shiu, Eric and Thakur, Rahul and Kim, Daehyun and Kuusela, Aki and Knies, Allan and Ranganathan, Parthasarathy and others},
  booktitle={Proceedings of the Twenty-Third International Conference on Architectural Support for Programming Languages and Operating Systems},
  pages={316--331},
  year={2018}
}

@inproceedings{shin2024processing,
  title={A Processing-using-Memory Architecture for Commodity {DRAM} Devices with Enhanced Compatibility and Reliability},
  author={Shin, Hoon and Park, Rihae and Lee, Jae W},
  booktitle={Proceedings of the 43rd IEEE/ACM International Conference on Computer-Aided Design},
  pages={1--10},
  year={2024}
}

@inproceedings{seshadri2013rowclone,
  title={{RowClone}: Fast and energy-efficient in-{DRAM} bulk data copy and initialization},
  author={Seshadri, Vivek and Kim, Yoongu and Fallin, Chris and Lee, Donghyuk and Ausavarungnirun, Rachata and Pekhimenko, Gennady and Luo, Yixin and Mutlu, Onur and Gibbons, Phillip B and Kozuch, Michael A and others},
  booktitle={Proceedings of the 46th Annual IEEE/ACM International Symposium on Microarchitecture},
  pages={185--197},
  year={2013}
}

@inproceedings{hajinazar2021simdram,
  title={{SIMDRAM}: A framework for bit-serial {SIMD} processing using {DRAM}},
  author={Hajinazar, Nastaran and Oliveira, Geraldo F and Gregorio, Sven and Ferreira, Jo{\~a}o Dinis and Ghiasi, Nika Mansouri and Patel, Minesh and Alser, Mohammed and Ghose, Saugata and G{\'o}mez-Luna, Juan and Mutlu, Onur},
  booktitle={Proceedings of the 26th ACM International Conference on Architectural Support for Programming Languages and Operating Systems},
  pages={329--345},
  year={2021}
}

@inproceedings{khan2024cinm,
  title={{CINM} ({Cinnamon}): A compilation infrastructure for heterogeneous compute in-memory and compute near-memory paradigms},
  author={Khan, Asif Ali and Farzaneh, Hamid and Friebel, Karl Friedrich Alexander and Fournier, Cl{\'e}ment and Chelini, Lorenzo and Castrillon, Jeronimo},
  booktitle={Proceedings of the 29th ACM International Conference on Architectural Support for Programming Languages and Operating Systems, Volume 4},
  pages={31--46},
  year={2024}
}

@inproceedings{yuksel2024functionally,
  title={Functionally-complete {Boolean} logic in real {DRAM} chips: Experimental characterization and analysis},
  author={Y{\"u}ksel, {\.I}smail Emir and Tu{\u{g}}rul, Yahya Can and Olgun, Ataberk and Bostanc{\i}, F Nisa and Ya{\u{g}}l{\i}k{\c{c}}{\i}, A Giray and Oliveira, Geraldo F and Luo, Haocong and G{\'o}mez-Luna, Juan and Sadrosadati, Mohammad and Mutlu, Onur},
  booktitle={2024 IEEE International Symposium on High-Performance Computer Architecture (HPCA)},
  pages={280--296},
  year={2024},
  organization={IEEE}
}

@article{khan2024landscape,
  title={The landscape of compute-near-memory and compute-in-memory: A research and commercial overview},
  author={Khan, Asif Ali and de Lima, Jo{\~a}o Paulo C and Farzaneh, Hamid and Castrillon, Jeronimo},
  journal={arXiv preprint arXiv:2401.14428},
  year={2024}
}

@inproceedings{seshadri2017ambit,
  title={Ambit: In-memory accelerator for bulk bitwise operations using commodity {DRAM} technology},
  author={Seshadri, Vivek and Lee, Donghyuk and Mullins, Thomas and Hassan, Hasan and Boroumand, Amirali and Kim, Jeremie and Kozuch, Michael A and Mutlu, Onur and Gibbons, Phillip B and Mowry, Todd C},
  booktitle={Proceedings of the 50th Annual IEEE/ACM International Symposium on Microarchitecture},
  pages={273--287},
  year={2017}
}

@article{olgun2022pidram,
  title={{PiDRAM}: A Holistic End-to-end {FPGA}-based Framework for Processing-in-{DRAM}},
  author={Olgun, Ataberk and G{\'o}mez-Luna, Juan and Kanellopoulos, Konstantinos and Salami, Behzad and Hassan, Hasan and Ergin, Oguz and Mutlu, Onur},
  journal={ACM Transactions on Architecture and Code Optimization},
  volume={20},
  number={1},
  pages={1--31},
  year={2022},
  publisher={ACM New York, NY}
}

@article{gem5,
      title={The {gem5} Simulator: Version 20.0+},
      journal={arXiv preprint arXiv:2007.03152},
      author={Jason Lowe-Power and Abdul Mutaal Ahmad and Ayaz Akram and Mohammad Alian and Rico Amslinger and Matteo Andreozzi and Adri{\`a} Armejach and Nils Asmussen and Brad Beckmann and Srikant Bharadwaj and Gabe Black and Gedare Bloom and Bobby R. Bruce and Daniel Rodrigues Carvalho and Jeronimo Castrillon and Lizhong Chen and Nicolas Derumigny and Stephan Diestelhorst and Wendy Elsasser and Carlos Escuin and Marjan Fariborz and Amin Farmahini-Farahani and Pouya Fotouhi and Ryan Gambord and Jayneel Gandhi and Dibakar Gope and Thomas Grass and Anthony Gutierrez and Bagus Hanindhito and Andreas Hansson and Swapnil Haria and Austin Harris and Timothy Hayes and Adrian Herrera and Matthew Horsnell and Syed Ali Raza Jafri and Radhika Jagtap and Hanhwi Jang and Reiley Jeyapaul and Timothy M. Jones and Matthias Jung and Subash Kannoth and Hamidreza Khaleghzadeh and Yuetsu Kodama and Tushar Krishna and Tommaso Marinelli and Christian Menard and Andrea Mondelli and Miquel Moreto and Tiago M{\"u}ck and Omar Naji and Krishnendra Nathella and Hoa Nguyen and Nikos Nikoleris and Lena E. Olson and Marc Orr and Binh Pham and Pablo Prieto and Trivikram Reddy and Alec Roelke and Mahyar Samani and Andreas Sandberg and Javier Setoain and Boris Shingarov and Matthew D. Sinclair and Tuan Ta and Rahul Thakur and Giacomo Travaglini and Michael Upton and Nilay Vaish and Ilias Vougioukas and William Wang and Zhengrong Wang and Norbert Wehn and Christian Weis and David A. Wood and Hongil Yoon and {\'E}der F. Zulian},
      year={2020},
      eprint={2007.03152},
      archivePrefix={arXiv}
}

@inproceedings{lee2024pim,
  title={{PIM-MMU}: A memory management unit for accelerating data transfers in commercial {PIM} systems},
  author={Lee, Dongjae and Hyun, Bongjoon and Kim, Taehun and Rhu, Minsoo},
  booktitle={2024 57th IEEE/ACM International Symposium on Microarchitecture (MICRO)},
  pages={627--642},
  year={2024},
  organization={IEEE}
}

@article{olgun2023drambender,
  author={Olgun, Ataberk and Hassan, Hasan and Ya{\u{g}}l{\i}k{\c{c}}{\i}, A. Giray and Tu{\u{g}}rul, Yahya Can and Orosa, Lois and Luo, Haocong and Patel, Minesh and Ergin, O{\u{g}}uz and Mutlu, Onur},
  title={{DRAM Bender}: An Extensible and Versatile {FPGA}-Based Infrastructure to Easily Test State-of-the-Art {DRAM} Chips},
  journal={IEEE Transactions on Computer-Aided Design of Integrated Circuits and Systems},
  volume={42},
  number={12},
  pages={5098--5112},
  year={2023},
  doi={10.1109/TCAD.2023.3282172}
}

@article{ramulatorpim,
  author={Kim, Yoongu and Yang, Weikun and Mutlu, Onur},
  title={{Ramulator}: A Fast and Extensible {DRAM} Simulator},
  journal={IEEE Computer Architecture Letters},
  volume={15},
  number={1},
  pages={45--49},
  year={2016},
  doi={10.1109/LCA.2015.2414456}
}

@misc{nvmainpim,
  author={{SU-JonesLab}},
  title={{NVMain-PIM}},
  year={2025},
  howpublished={\url{https://github.com/SU-JonesLab/NVMain-PIM}}
}

\end{document}